\documentstyle[aps]{revtex}

\begin{document}

\title{Existence, uniqueness and other properties of the BCT (minimal
strain lapse and shift) gauge}

\author{David Garfinkle}
\address{Department of Physics, Oakland University, Rochester, MI
48309, USA}
\author{Carsten Gundlach}
\address{Faculty of Mathematical Studies, University of Southampton,
Southampton SO17 1BJ, United Kingdom}
\author{James Isenberg}
\address{Department of Mathematics, University of Oregon, Eugene, OR
97403, USA}
\author{Niall \'OMurchadha}
\address{Physics Department, University College, Cork, Ireland}

\date{31.3.00}

\maketitle

\begin{abstract}
Brady, Creighton and Thorne have proposed a choice of the lapse and
shift for numerical evolutions in general relativity that extremizes a
measure of the rate of change of the three-metric (BCT gauge). We
investigate existence and uniqueness of this gauge, and comment on its
use in numerical time evolutions.
\end{abstract}

\section{The BCT gauge}

Cauchy data for general relativity consist of a three-metric $g_{ab}$
and extrinsic curvature $K_{ab}$ specified on a three-manifold
(``slice'') $\Sigma$. The Cauchy data determine the four-dimensional
spacetime (locally) as a geometric object, but without fixing a
coordinate system. When the spacetime is computed numerically as a
sequence of spacelike slices $\Sigma(t)$, the coordinates may be fixed
incrementally by specifying a lapse $\alpha$ and shift vector
$\beta^a$ on each new slice -- a gauge choice. In one class of gauge
choices, $\alpha$ and $\beta^a$ on a slice are determined by $g_{ab}$
and $K_{ab}$ on the same slice, i.e.,
\begin{equation}
(\Sigma, g_{ab}, K_{ab}) \to (\alpha,\beta^a).
\end{equation}

Recently, Brady, Creighton and Thorne \cite{BCT} have proposed a new
gauge choice of this class in which $\alpha$ and $\beta^a$ are
determined as a solution of the coupled equations
\begin{equation}
\label{BCT1}
K^{ab}F_{ab} = 0, \qquad D^a F_{ab}= 0,
\end{equation}
where $D_a$ is the covariant derivative with respect to the
three-metric $g_{ab}$, and
\begin{equation}
F_{ab}(\alpha,\beta) \equiv L_{ab}(\beta)-2\alpha K_{ab}, \qquad
L_{ab}(\beta) \equiv D_a\beta_b + D_b\beta_a
\end{equation}
These two equations arise when one varies the action principle
\begin{equation}
I = \int F^{ab} F_{ab} \, \sqrt{g}\, d^3 x
\end{equation}
with respect to $\alpha$ and $\beta^b$, respectively. In a
four-dimensional context, $F_{ab}$ is the time derivative of the
three-metric induced on the time slicing with lapse $\alpha$ and shift
$\beta^a$:
\begin{equation}
\dot g_{ab} \equiv {\cal L}_t g_{ab} = F_{ab}(\alpha,\beta), \qquad
t^a \equiv \alpha n^a +\beta^a,
\end{equation}
where $n^a$ is the unit normal vector of the slice $\Sigma$ when it is
embedded into a spacetime. Equations (\ref{BCT1}) are therefore also
called the minimal strain lapse equation and minimal strain shift
equation, where $I$ is the ``strain'' that is being extremized.

The motivation for considering these equations is that a good gauge
choice should have the property (among others) to be compatible with
approximate Killing vectors, in the sense that if an approximate
Killing vector exists, the spacetime metric in that gauge should
evolve as slowly as possible. The inspiral phase of a binary black
hole system in a spherical orbit, for example, has an approximate
Killing vector, and one would like to be able to find (corotating)
coordinates in which the spacetime metric evolves on the timescale in
which the orbit shrinks through the emission of gravitational waves,
rather than on the much shorter orbital period timescale.

The first of the two equations (\ref{BCT1}) can be solved
algebraically for $\alpha$,
\begin{equation}
\label{BCTlapse}
\alpha = {K^{cd} L_{cd}(\beta)\over K^{ef} K_{ef} } ,
\end{equation}
and the result substituted into the second equation. One obtains a
vector linear second-order differential equation for the vector
$\beta$ alone:
\begin{equation}
\label{BCT}
D^a H_{ab}(\beta) = 0, \qquad H_{ab}(\beta) \equiv L_{ab}(\beta) - 2
{K^{cd} L_{cd}(\beta)\over K^{ef} K_{ef} } K_{ab}.
\end{equation}
We shall call this single vector equation the BCT equation, and in the
following we shall consider the lapse $\alpha$ as a dependent quantity
determined by (\ref{BCTlapse}). Equation (\ref{BCT}) can be obtained
from the action principle
\begin{equation}
J = \int H^{ab} H_{ab} \, \sqrt{g}\, d^3 x,
\end{equation}
which is obtained by substituting the lapse (\ref{BCTlapse}) into the
action principle $I$.

\section{Existence and uniqueness}

If one is to use the BCT gauge, it is important to know the answer to
the following question: For which choices of data $(U,g_{ab},K_{ab})$
on a region $U$ of $\Sigma$ with boundary $\partial U$ and for which
sets of boundary conditions for $\beta$ can one solve the BCT equation
(\ref{BCT})? In this brief note, we show that for some choices of data
there is a unique solution with specified Dirichlet boundary
conditions, while for others, there are many solutions with those
boundary conditions.

The issues of existence and uniqueness of the BCT gauge have been
previously considered by Gon\c calves \cite{Goncalves}. Gon\c calves
shows that the differential operator defined by (\ref{BCT}) is
strongly elliptic if and only if ${K_a}^b$, considered as a map on the
tangent space of $\Sigma$, has at most one vanishing eigenvalue. (Let
the principal part of the differential operator acting on $\beta^a$ be
${{M^a}_b}^{cd} D_c D_d \beta^b$. The operator is then defined to be
strongly elliptic if ${{M^a}_b}^{cd}$ is positive definite with
respect both to the two indices that slot into derivatives and the
index pair that shuffles the vector index on $\beta$.) Generally,
operator ellipticity -- strong or otherwise -- is not enough to
determine whether a boundary value problem admits a solution. However,
using the fact that the equation of interest (\ref{BCT}) is of
divergence form, Gon\c calves does obtain local results. In this
paper, using the Fredholm alternative, we obtain stronger, global,
results.

Before deriving the condition for existence of the BCT gauge, we wish
to turn the boundary value problem for Equation (\ref{BCT}) into one
with homogeneous boundary conditions. Let us write Equation
(\ref{BCT}) in operator form as $O(\beta) = 0$, and the corresponding
Dirichlet problem as
\begin{equation}
\label{Dirichlet}
O(\beta) = 0, \quad \beta|_{\partial U} = f
\end{equation}
for some given continuous vector-valued function $f$. Presuming that
the region $U$ and its boundary are well-behaved, we may always extend
$f$ to a function $F:U\to R^3$ which is $C^2$ on $U$ and continuous on
$U\cup\partial U$. Then if we find a solution $\xi$ of the boundary
value problem
\begin{equation}
\label{Dirichlet_hom}
O(\xi) = - O(F), \quad \xi|_{\partial U} = 0,
\end{equation}
and if we set $\beta=\xi+F$, we have $\beta$ satisfying the boundary
value problem (\ref{Dirichlet}). We may now focus on the discussion of
the boundary value problem (\ref{Dirichlet_hom}) (for arbitrary $F$).

To be able to use Fredholm ideas to study the boundary value problem
(\ref{Dirichlet_hom}), one first needs to establish that
(\ref{Dirichlet_hom}) defines a Fredholm map. As shown in propositions
11.10 and 11.16 of \cite{Fredholm}, this fact follows from the strong
ellipticity of $O$. We also need the following:

{\bf Lemma}: For $\beta$ satisfying the Dirichlet condition
$\beta|_{\partial U}=0$, $O$ is a self-adjoint operator.

{\bf Proof}: We consider the quantity
\begin{equation}
\int_U \gamma \, O(\beta) \equiv \int_U \gamma^a D^b H_{ab}(\beta)
\end{equation}
with $H_{ab}$ from Equation (\ref{BCT}). Integrating by parts, and
using $\gamma|_{\partial U}=0$, we obtain
\begin{equation}
\int_U \gamma \, O(\beta) = - \int_U D^b \gamma^a H_{ab}(\beta) = -
\int_U H^{ab}(\gamma) D_a\beta_b.
\end{equation}
Integrating by parts again, and using $\beta|_{\partial U}=0$, we have
\begin{equation}
\int_U \gamma \, O(\beta) = \int_U D_a H^{ab}(\gamma) \beta_b = \int_U
\beta \, O(\gamma) .
\end{equation}
Hence $O$ is self-adjoint on functions $\beta$ that vanish on the
boundary. (As the boundary term in the integration by parts is of the
form $\int_{\partial U} s^a \beta^b H_{ab}(\gamma)$, where $s^a$ is
the unit normal vector to $\partial U$, the operator $O$ is also
self-adjoint on the space of functions $\gamma$ that satisfy $s^a
H_{ab}(\gamma)=0$ on the boundary. This boundary condition is the
mathematical analog of the Neumann boundary condition for the Laplace
equation, but it contains additional terms that make its physical
meaning unclear. Therefore we do not consider it here.)

Since $O$ is self-adjoint, and since it defines a Fredholm map, the
Fredholm Alternative \cite{Fredholm} specifies a clear condition for
existence of solutions. Stated for the operator $O$, one has

{\bf Proposition} (Fredholm Alternative): Fix $(U,g_{ab},K_{ab})$. The
Dirichlet boundary value problem (\ref{Dirichlet_hom}) has a unique
solution for every choice of $O(F)$ if and only if the only vector
function satisfying
\begin{equation}
\label{kernel_general}
O(\xi)=0, \quad \xi|_{\partial U}=0
\end{equation}
is $\xi^a=0$. (That is, the kernel of $O$ is empty.)  For a given
$O(F)$, the boundary value problem (\ref{Dirichlet_hom}) has a
solution so long as $O(F)$ satisfies
\begin{equation}
\int_U \xi \, O(F) = 0
\end{equation}
for every $\xi$ which satisfies (\ref{kernel_general}). (The kernel of
$O$ is orthogonal to the source.)

We base our existence results for the BCT gauge on the Fredholm
Alternative. So we are led to consider solutions $\xi$ of
(\ref{kernel_general}), or elements of the kernel of $O$. Given such a
solution, we consider the quantity
\begin{equation}
\int_U \xi \, O(\xi) = -{1\over 2} \int_U L^{ab}(\xi) H_{ab}(\xi) =
-{1\over 2} \int_U H^{ab}(\xi) H_{ab}(\xi),
\end{equation}
where the first equality follows from integration by parts ($\xi$
vanishes on $\partial U$) and the second equality follows from $K^{ab}
H_{ab}=0$. $\xi$ therefore satisfies (\ref{kernel_general}) if and
only if
\begin{equation}
\label{kernel_BCT}
H_{ab}(\xi) = 0, \quad \xi|_{\partial U}=0.
\end{equation}
Thus the kernel of $O$ consists of solutions of this boundary value
problem.

For a given set of initial data $(U,g_{ab},K_{ab})$, the Fredholm
Alternative asks that we determine the kernel of $O$. If the kernel is
empty, then the Dirichlet boundary value problem
(\ref{Dirichlet_hom}), and hence (\ref{Dirichlet}), admits a unique
solution.

If the kernel is not empty, with non-trivial elements $\xi$, then to
determine the solubility of the boundary value problem
(\ref{Dirichlet_hom}) we need to consider $\int_U \xi \, O(F)$.
Integrating by parts, we have
\begin{equation}
\int_U \xi \, O(F) = - {1\over 2} \int_U L(\xi) H(F) = - {1\over 2}
\int_U H(\xi) L(F),
\end{equation}
but $H(\xi)=0$ by assumption. So this always vanishes, and a solution
always exists in this case, too.  It is determined, however, only up
to the addition of any element of the kernel.

We conclude that a solution of the BCT gauge exists whenever $O$ is
strongly elliptic, but if the kernel (\ref{kernel_BCT}) is not empty,
the solution is not unique.

\section{Examples for uniqueness and non-uniqueness}

It is useful to note that there are choices of data
$(U,g_{ab},K_{ab})$ for which each of these two cases hold. For the
first case, where the kernel is empty, we consider data with
$K_{ab}=\rho g_{ab}$, with $\rho$ nowhere vanishing. The kernel
equation (\ref{kernel_BCT}) then becomes
\begin{equation}
D_a\xi_b + D_b\xi_a - {2\over 3} g_{ab} D^c\xi_c = 0,
\end{equation}
This is the equation for a conformal Killing vector. If we choose a
metric $g_{ab}$ that does not admit a conformal Killing vector that
vanishes on the boundary, we have constructed data that give rise to
an empty kernel, and therefore a unique BCT gauge for a given choice
of the boundary data.

For the other case, we consider a slice through a spacetime that has a
timelike Killing vector. We choose the lapse and shift so that $t^a$
is the Killing vector. The time derivative of the three-metric then
vanishes, and consequently $F_{ab}(\alpha,\beta)=0$. From this it
follows that $H_{ab}(\beta)=0$. In order to obtain $\beta|_{\partial
U}=0$, we choose the slice so that it is normal to the Killing vector
in $\partial B$ but not in $B$. If in such a situation one tries to
solve the BCT equation with a boundary where the slice is
approximately normal to the Killing vector, the numerical problem
might become badly conditioned.

As a concrete example, we consider a spherically symmetric slice in
flat spacetime. Let $(t,r,\theta,\varphi)$ be the standard spherically
coordinates on Minkowski spacetime, and let the slice be given by
$t=T(r)$. As coordinates intrinsic to the slice we use
$(r,\theta,\varphi)$ induced by the spacetime coordinates of the same
name. The normal vector to the slice, induced 3-metric and extrinsic
curvature are then given by
\begin{eqnarray}
n^r &=& {T'\over \sqrt{1-T'^2}}, \quad n^t = {1\over \sqrt{1-T'^2}},
\\ g_{rr} &=& 1-T'^2, \quad g_{\theta\theta} = r^2, \quad
g_{\varphi\varphi}=r^2\sin^2\theta, \\ K_{rr} &=& - {T''\over
\sqrt{1-T'^2}}, \quad K_{\theta\theta} = -{rT' \over \sqrt{1-T'^2}},
\quad K_{\varphi\varphi} = \sin^2\theta K_{\theta\theta}.
\end{eqnarray}
The normal vector is parallel to the Killing vector $\partial/\partial
t$ where $T'(r)=0$, say at $r=r_0$. We choose $U$ to be the ball $r\le
r_0$. The desired element of the kernel of $O$ is then
\begin{equation}
\alpha = \sqrt{1-T'^2}, \quad \beta^r = - T',
\end{equation}
up to an overall constant factor.

The potential difficulty with non-uniqueness can be avoided by an
appropriate choice of slice and boundary.  In particular, for the
choice of slice and boundary proposed in \cite{BCT} for the binary
black hole problem, the shift is nowhere small, so that the Killing
vector is nowhere normal to the slice. Near the black hole excision
boundary, the slicing is of the Painlev\'e-G\"ullstrand (or
Kerr-Schild) type, with a large radial shift, while at the outer
boundary the coordinates are corotating, with a large
$\partial/\partial\phi$ shift component.

\section{Use of the BCT gauge and other gauges in time evolution}

Consider the evolution of initial data $(g_{ab},K_{ab})$. The
evolution equation for $K_{ab}$ is of the form
\begin{equation}
\dot K_{ab} = - D_a D_b \alpha + \hbox{other terms}.
\end{equation}
The time derivative of $K_{ab}$ contains the second spatial derivative
of the lapse. If one uses the BCT gauge to determine $\alpha$ and
$\beta$ from $g_{ab}$ and $K_{ab}$, the evolution equation for
$K_{ab}$ becomes
\begin{equation}
\label{K_evolution}
\dot K_{ab} = - \left[ {L^{cd}(\beta)\over K_{ef}K^{ef}} - 2
{L^{mn}(\beta)K_{mn} K^{cd} \over (K_{ef}K^{ef})^2} \right] D_a D_b
K_{cd} + \hbox{other terms}.
\end{equation}
This means that if $K_{ab}$ is initially in a function space of finite
differentiability, time evolution takes it out of that space -- it
``loses two derivatives''. ($\beta^a$ itself appears to gain two
derivatives because it is the solution of an elliptic equation, but
this does not affect the argument.)

While this is a technical obstacle to proving existence and uniqueness
of solutions to the Einstein equations in the BCT gauge, it also hints
at the possible existence of a practical problem for the use of the
BCT gauge in numerical evolution.  Roughly speaking, one would expect
numerical noise to be amplified during time evolution, whereas maximal
slicing with zero shift, for example, is empirically known to dampen
numerical noise. In the toy model equation $\dot u=\kappa u_{xx}$,
noise is damped for $\kappa>0$ (heat equation) and increases for
$\kappa<0$ (heat equation run backwards), so that one only needs to
choose the correct sign of $\kappa$. In contrast, Equation
(\ref{K_evolution}) is a nonlinear tensor equation. It appears
plausible that some of the eigenvalues of its linearization around
some backgrounds $(g_{ab},K_{ab})$ correspond to negative $\kappa$,
thus leading to the growth of linear noise. Corresponding nonlinear
instabilities may also exist. We have not attempted to investigate
this question.

Here we would like to draw attention to an alternative gauge choice.
Consider the following coupled equations for the lapse and shift:
\begin{eqnarray}
{D_b} {D^b} \beta^a + D_b D^a \beta^b - 2 D_b(\alpha K^{ab}) - {n\over
3} D^a(2D_b\beta^b - 2\alpha K) & = & 0, \\
-D_aD^a\alpha+\left[{}^{(3)}R+K^2 +{1\over2}(\tau-3\rho)\right]\alpha
+ \beta^aD_aK & = & 0.
\end{eqnarray}
Here $\tau$ and $\rho$ are matter terms, namely the trace of the
(three-dimensional) stress tensor and the energy density. $n$ is a
constant that is either zero or one. These equations for $\alpha$ and
$\beta$ are elliptic, but they are not self-adjoint. Existence and
uniqueness of solutions will be considered elsewhere \cite{Niall}. As
the equations are elliptic in both $\alpha$ and $\beta$, $(\dot
g_{ab},\dot K_{ab})$ do not lose differentiability compared to the
Cauchy data $(g_{ab},K_{ab})$.

The first, vector, equation is $D_b F^{ab}=0$ for $n=0$, and is $D_b
(F^{ab}-{1/3}g^{ab} {F^c}_c)=0$ for $n=1$. These two shift conditions
were suggested by Smarr and York in their classic paper on coordinate
conditions \cite{SmarrYork} under the names ``minimal strain shift''
and ``minimal distortion shift''. The second, scalar, equation is
$\dot K=0$. For $K=0$ it reduces to maximal slicing. Here, however, we
do not assume that $K$ is zero, nor that it is constant in
space. $\dot K=0$ slicing, and its combination with minimal distortion
shift, was suggested by Smarr and York, and they also discuss its
Killing vector-tracking property. Their subsequent discussion focuses
on maximal ($K=0$) slicing with minimal distortion shift. This more
restricted gauge choice is now commonly associated with the name
``Smarr-York'' (SY) gauge, while the more general gauge with $K\ne 0$
seems to have been forgotten. The desirability of Killing vector
tracking was later rediscovered in \cite{BCT} and \cite {symmcoord},
and the gauge discussed here was rediscovered as ``generalized
Smarr-York'' (GSY) gauge in \cite{symmcoord}.

Here we want to point out that the GSY gauge has the desirable
properties of the BCT gauge -- it tracks Killing vectors and it admits
generic Cauchy data -- without having the ``loss of derivatives''
property and the possible problem when the slice is normal to the
Killing vector at the boundary. In a direct numerical comparison of
the BCT and GSY gauges in spherical symmetry, Garfinkle and Gundlach
\cite{symmcoord} find that the GSY gauge stably tracks Killing
vectors, but are unable to obtain a stable time evolution with the BCT
gauge, and this may be due to the ``loss of derivatives'' property.
Shibata \cite{Shibata} reports good experiences with an approximate
implementation of maximal slicing with minimal strain shift in 3D
evolutions of a neutron star binary. Independent numerical work, and
tests in 3D, are required to decide if one of the gauges is more
suitable in practice than the other.

\acknowledgments

This work was carried out during a mini-program on ``Colliding Black
Holes'' at the Institute for Theoretical Physics, Santa Barbara, where
it was supported by NSF grant PHY-9407194. DG is also supported NSF
grant PHY-9722039 to Oakland University, and JI by grant PHY-9800732
to the University of Oregon.

\end{document}